\documentstyle[thmsa,sw20aip]{article}

\input{tcilatex}
\input tcilatex
\QQQ{Language}{
American English
}

\begin{document}

{\bf Electromagnetic Waves in the Vacuum with Torsion and Spin.}\\ 

\qquad \qquad R. M. Kiehn

\qquad \qquad {\it Physics Department, University of Houston,}\thinspace 
{\it \ Houston, TX, 77004}

\begin{quotation}
{\bf \ }Exact radiative wave solutions to the classical homogeneous Maxwell
equations in the vacuum have been found that are not transverse, exhibit
both torsion and spin, and for which the second Poincare invariant ${\bf E}%
\circ {\bf B}\neq 0.$ \ Two four component rank 3 tensors of spin current
and torsion are constructed on topological grounds. \ The divergence of each
pseudo vector generates the Poincare invariants of the electromagnetic
system. \ \\ 

PACS numbers 03.50.De, 41.10.Hv\\
\end{quotation}

{\bf 0. Introduction}

\qquad In section 1 the domain of classical electromagnetism is defined in
terms of four vector fields ${\bf D,E,B,H,}$ and the vector and scalar
potentials $\{{\bf A},\phi \}{\bf .}$ \ In section 2 prior attempts to find
time dependent wave solutions with non-zero Poincare invariants are
discussed briefly, with special emphasis placed upon Ranada's use of the
Hopf map. \ In section 3, several time dependent closed form solutions are
presented that have ${\bf E}\circ {\bf B}\neq 0$. \ \\ 

{\bf 1. The Domain of Classical Electromagnetism}

\qquad In terms of the notation and the language of Sommerfeld and Stratton
[1], the classic definition of an electromagnetic system is a domain of
space-time independent variables, $\{x,y,z,t\},$ which supports both the
Maxwell-Faraday equations,

\begin{equation}
curl\,\,{\bf E}+\partial {\bf B}/\partial t=0,\,\,\,\,\,\,\,\,\,\,div\,\,%
{\bf B}=0,  \tag*{(1.1)}
\end{equation}
and the Maxwell-Ampere equations,

\begin{equation}
curl\,\,{\bf H}-\partial {\bf D}/\partial t={\bf J},\,\,\,\,\,\,\,\,\,\,div%
\,\,{\bf D}=\rho .  \tag*{(1.2)}
\end{equation}

For the Lorentz vacuum state, the charge-current densities are subsumed to
be zero $\lbrack {\bf J}$, $\rho \rbrack =0$ and the field excitations, $%
{\bf D}$ and ${\bf H}$, are linearly connected to the field intensities, $%
{\bf E}$ and ${\bf B}$, \ by means of the homogeneous and isotropic
constitutive relations ${\bf D}=\varepsilon {\bf E}\,,\,\,\,\,\,{\bf B}=\mu 
{\bf H.}$ \ It is further subsumed that the classic Maxwell electromagnetic
system is constrained by the \ statement that the field intensities are
deducible from a system of twice differentiable potentials, $\lbrack {\bf A}%
,\phi \rbrack $:

\begin{equation}
{\bf B}=curl\,{\bf \,A,\,\,\,\,\,\,\,\,E}=-grad\,\,\phi -\partial {\bf A}%
/\partial t.  \tag*{(1.3)}
\end{equation}
This constraint topologically implies that domains that support non-zero
values for the covariant field intensities, ${\bf E}$ and ${\bf B,}$ can 
{\it not} be compact domains without a boundary, unless the domain has Euler
characteristic zero. \ The only two exceptions are therefore the Torus and
the Klein bottle.

\qquad Besides the charge current 4-vector density, $[{\bf J},\rho ],$ whose
integral over any closed 3 dimensional manifold is a deformation invariant
of the Maxwell system, there exist two other algebraic combinations of the
fields and potentials that can lead to similar topological quantities. \
These objects are the rank 3 Spin (pseudo) vector, or current [2], defined
in component form as

\begin{equation}
{\bf S}_4=[{\bf A\times H}+{\bf D}\phi ,{\bf A\circ D}]\equiv [{\bf S,}%
\sigma {\bf ]},  \tag*{(1.4)}
\end{equation}
and the rank 3 Torsion (pseudo) vector [3]\ defined in component form as

\begin{equation}
{\bf T}_4=[{\bf E\times A}+{\bf B}\phi ,{\bf A\circ B}]\equiv [{\bf T,}h{\bf %
]}.  \tag*{(1.5)}
\end{equation}
Note that the classical helicity, $h={\bf A\circ B,\,}$forms only the fourth
component of this third rank tensor. The derivation of these 4-component
tensor fields of rank 3 and their topological implications are developed in
more detail elsewhere.[4]\ \ The 4-divergence of these 4-component vectors
leads to the Poincare projective invariants of the Maxwell system:

\begin{eqnarray}
Poincare\,Invariant\,\,1 &=&div_3({\bf A\times H}+{\bf D}\phi )+\partial (%
{\bf A\circ D)}/\partial t  \tag*{(1.6)} \\
&=&({\bf B\circ H-D\circ E)-(A\circ J}-\rho \phi )  \nonumber
\end{eqnarray}

\begin{eqnarray}
Poincare\,\,Invariant\,\,2 &=&div_3({\bf E\times A}+{\bf B}\phi )+\partial (%
{\bf A\circ B)}/\partial t  \tag*{(1.7)} \\
&=&-2{\bf E\circ B}  \nonumber
\end{eqnarray}

When the Spin vector is non-zero, and its 4-divergence (the first Poincare
invariant) vanishes, integrals over closed three manifolds of the Spin 4
vector lead a topological property equivalent to a deRham period integral
[5]:

\begin{equation}
Spin=\tiiint_{closed}\{S^xdy\symbol{94}dz\symbol{94}dt-S^ydx\symbol{94}dz%
\symbol{94}dt+S^zdx\symbol{94}dy\symbol{94}dt-\sigma dx\symbol{94}dy\symbol{%
94}dz\}.  \tag*{(1.8)}
\end{equation}
This closed integral is a deformation invariant of any evolutionary process
that can be described by a singly parameterized vector field,$\,\beta {\bf V,%
}$ independent of the choice of parameterization, $\beta $, for the Lie
derivative of the Spin integral vanishes:

\begin{equation}
L_{(\beta {\bf V)}}\,Spin=0.  \tag*{(1.9)}
\end{equation}
When the associated Poincare invariant vanishes, the values of the Spin
integral form rational ratios. \ Similar statements hold for the closed
integrals of the Torsion vector.\\ 

{\bf 2. Earlier Work}

\qquad In earlier articles, Chu and Ohkawa [6]\ developed a standing wave
example that led Khare and Pradhan [7]\ to construct a free space
electromagnetic wave which had non-zero Poincare Invariants. \ Braunstein
[8]\ mentioned that these developments were technically flawed and further
argued that the existence of a bonafide (spatially bounded) electromagnetic
wave in free space with non-zero Poincare invariants was impossible.

\qquad The solution counter examples to Braunstein's claim, as given in
section 3 below, were inspired by the work of \ Ranada [9]\ who investigated
the applications of the Hopf map to the problem of finding knotted solutions
to the Maxwell equations. \ Recall that the Hopf map can be written as the
common constraint on the map$\,\,\Phi $ from $R4(x,y,z,s)$ to $R3(X,Y,Z)$
given by the expressions:

\begin{equation}
\lbrack X,Y,Z]=[2(ys-xz)\,,-2(yz+xs)\,\,,-(z^2+s^2)+(x^2+y^2)]  \tag*{(2.1)}
\end{equation}
\qquad From another point of view, the\ Hopf map defines a family of
cyclides in $\{x,y,z\}$ parameterized by $s$

\begin{equation}
Hopf\,\,\ Map\,Cyclide\,\,\,\,\,\,\,\,\,\,\,\,\,\,\,\,r^4+(2s^2-1)r^2+s^4=0,
\tag*{(2.2)}
\end{equation}
and where $r^2=x^2+y^2+z^2.\,$\ \ A picture of the Hopf cyclide can be seen
in reference [10].

\qquad Ranada suggested the 4-potential (based on the Hopf map for $s=1$)

\begin{equation}
{\bf A}=[y,-x,-1](2/\pi )/\lambda ^4\,\,,\,\,\,\,\,\,\phi =0/\lambda
^4,\,\,\,\,\,where\,\,\lambda ^2=1+x^2+y^2+z^2,  \tag*{(2.3)}
\end{equation}
which will generate the fields

\begin{equation}
{\bf E}=[0,0,0]\,\,\,\,\,\,\,\,\,\,\,\,\,\,\,\,\,{\bf B}%
=[-2(y+zx),+2(x-yz),+(-1+x^2+y^2-z^2)](4/\pi )/\lambda ^6.  \tag*{(2.4)}
\end{equation}
\ \ Note that the components of the induced ${\bf B}$ field are precisely \
the coefficients of the Hopf Map (to within a factor).\ \ \ Ranada discusses
the knottedness of the magnetic field lines of such solutions to the
Maxwell-Faraday equations, which have finite helicity, but zero second
Poincare invariant.

\begin{equation}
h={\bf A\circ B}=-8s/\pi ^2\lambda
^8\,\,\,,\,\,\,\,\,\,\,\,\,\,\,\,\,\,\,\,\,\,\,\,{\bf E\circ B}=0\,. 
\tag*{(2.5)}
\end{equation}
Unfortunately, the Ranada 4-potential does not satisfy the Maxwell-Ampere
equation for the vacuum with a zero charge current 4-vector, and therefore
is not a suitable vacuum solution.

Consider a modification of the Hopf map by substituting $s\Rightarrow ict$
to yield the modified time dependent potentials:

\begin{equation}
{\bf A}=[y,-x,+ict](2/\pi )/\lambda ^4\,\,,\,\,\,\,\,\,\phi =icz/\lambda
^4,\,\,\,\,\,where\,\,\lambda ^2=-(ct)^2+x^2+y^2+z^2.  \tag*{(2.6)}
\end{equation}
Such potentials lead to complex ${\bf E}$ and ${\bf B}$ fields that indeed
satisfy (subject to the phase condition $\varepsilon \mu c^2=1)$ the zero
charge current criteria for a vacuum solution, and the vector wave equation.
\ However, the second Poincare invariant is imaginary and the Poynting
vector vanishes for such solutions. \ The Spin vector, on the other hand, is
real and has non-zero divergence. \ \\

{\bf 3. \ Example Radiative Vacuum Solutions for which }${\bf E}\circ {\bf B}%
\neq 0.$

The modifications of the Hopf map further suggest consideration of the
system of potentials given by the equations

\begin{equation}
{\bf A}=[+y,-x,-ct]/\lambda ^4\,\,,\,\,\,\,\,\,\phi =cz/\lambda
^4,\,\,\,\,\,where\,\,\lambda ^2=-c^2t^2+x^2+y^2+z^2.  \tag*{(3.1)}
\end{equation}
which yield the real field intensities,

\begin{equation}
{\bf E}=[-2(cty-xz),+2(ctx+yz),-(c^2t^2+x^2+y^2-z^2)]2c/\lambda ^6 
\tag*{(3.2)}
\end{equation}
and

\begin{equation}
{\bf B}=[-2(cty+xz),+2(ctx-yz),+(c^2t^2+x^2+y^2-z^2)]2/\lambda ^6. 
\tag*{(3.3)}
\end{equation}
Subject to the dispersion relation, $\varepsilon \mu c^2=1$ and the Lorentz
constitutive conditions, these time dependent wave functions satisfy the
homogeneous Maxwell equations without charge currents, and are therefore
acceptable vacuum solutions. The extensive algebra involved in these and
other computations in this article were checked with a Maple symbolic
mathematics program [11].

\qquad The Spin current density for this first non-transverse wave example
is evaluated as:

\begin{equation}
{\bf S}_4=[x(3\lambda ^2-4y^2-4x^2),y(3\lambda ^2-4y^2-4x^2),z(\lambda
^2-4y^2-4x^2),t(\lambda ^2-4y^2-4x^2)](2/\mu )/\lambda ^{10},  \tag*{(3.4)}
\end{equation}
and has zero divergence. \ The Torsion current may be evaluated as

\begin{equation}
{\bf T}_4=-[x,y,z,t]2c/\lambda ^8.  \tag*{(3.5)}
\end{equation}
and has a non-zero divergence equal to the second Poincare invariant

\begin{equation}
Poincare\,\,2=-2{\bf E}\circ {\bf B=+}8c/\lambda ^8.  \tag*{(3.6)}
\end{equation}
\qquad As the first Poincare invariant is zero it is possible to construct a
deformation invariant in terms of the deRham period integral of the Spin
current 4 vector over a closed 3 dimensional submanifold.

\qquad It is to be noted that the example solution given above is but one of
a class of vacuum wave solutions that have similar non transverse
properties. \ As a second example, consider the fields that can be
constructed from the potentials,

\begin{equation}
{\bf A}=[+ct,-z,+y]/\lambda ^4\,\,,\,\,\,\,\,\,\phi =cx/\lambda
^4,\,\,\,\,\,where\,\,\lambda ^2=-c^2t^2+x^2+y^2+z^2.  \tag*{(3.7)}
\end{equation}
These potentials will generate the field intensities

\begin{equation}
{\bf E}=[+(-c^2t^2+x^2-y^2-z^2),+2(ctz+yx),-2(cty-zx)]2c/\lambda ^6 
\tag*{(3.8)}
\end{equation}
and

\begin{equation}
{\bf B}=[+(-c^2t^2+x^2-y^2-z^2),+2(-ctz+yx),+2(cty+zx)]2/\lambda ^6. 
\tag*{(3.9)}
\end{equation}
As before, these fields satisfy the Maxwell-Faraday equations, and the
associated excitations satisfy the Maxwell-Ampere equations without
producing a charge current 4-vector. \ However, it follows by direct
computation that the second Poincare invariant, and the Torsion 4-vector are
of opposite signs to the values computed for the first example:

\begin{equation}
{\bf T}_4=+[x,y,z,t]2c/\lambda ^8\,,\,\,\,\,\,\,\,\,\,\,\,\,\,\,\,\,\,-2{\bf %
E}\circ {\bf B}=-8c/\lambda ^8\,\,\,.  \tag*{(3.10)}
\end{equation}

When the two examples are combined by addition (or subtraction), the
resulting wave is transverse magnetic (in the topological sense that ${\bf %
A\circ B}=0$). \ Not only does the second Poincare invariant vanish under
superposition, but so also does the Torsion 4 vector. \ Conversely, the
examples above show that there can exist transverse magnetic waves which can
be decomposed into two non-transverse waves. \ \thinspace \ A \ notable
feature of the superposed solutions is that the Spin 4 vector current does
not vanish, hence the example superposition is a wave that is not transverse
electric (${\bf A\circ D}\neq 0)$. \ For the examples presented above and
their superposition, the first Poincare invariant vanishes, which implies
that the Spin integral remains a conserved topological quantity for the
superposition, with values proportional to the integers. The Spin current
density for the combined examples is given by the formula:

\begin{eqnarray}
{\bf S}_4 &=&[-2x(y+ct)^2,(y+ct)(x^2-y^2+z^2-2cty-c^2t^2),-2z(y+ct)^2, 
\nonumber \\
&&-(y+ct)(x^2+y^2+z^2+2cty+c^2t^2)](4/\mu )/\lambda ^{10},  \tag*{(3.11)}
\end{eqnarray}
while the Torsion current is a zero vector

\begin{equation}
{\bf T}_4=[0,0,0,0].  \tag*{(3.12)}
\end{equation}

In addition, for the superposed example, the spatial components of the
Poynting vector are equal to the Spin current density vector multiplied by \ 
$\gamma $, such that

\begin{equation}
{\bf E\times H}=\gamma \,\,{\bf S},\,\,\,\,\,\ with\,\,\,\gamma
=-(x^2+y^2+z^2+2cty+c^2t^2)/2c(y+ct)\lambda ^2.  \tag*{(3.13)}
\end{equation}
These results seem to give classical credence to the Planck assumption that
vacuum state of \ Maxwell's electrodynamics supports quantized angular
momentum, and that the energy flux must come in multiples of the spin
quanta. \ In other words, these combined solutions to classical
electrodynamics have some of the qualities of the photon.\\

{\bf References}

{\small 1. A. Sommerfeld, Electrodynamics (Academic, New York, 1952). \
J.A.Stratton, Electromagnetic Theory McGraw Hill N.Y. 1941 \ }

{\small Sommerfeld carefully distinguishes between intensities and
excitations on thermodynamic grounds.}

{\small 2. R.M. Kiehn, and J.F. Pierce, Phys. Fluids {\bf 12}, 1971 (1969)}

{\small 3. \ \ R. M.Kiehn, Int. Journ. Mod Phys {\bf 5}, 10, 1779 (1991)}

{\small 4. \ See http://www.uh.edu/\symbol{126}rkiehn/pdf/helicity.pdf for a
preprint}

5. \ {\small R. M.}\ {\small Kiehn, J. of Math Phys {\bf 18}, no. 4, 614
(1977)}

{\small 6. C. Chu and T. Ohkawa, Phys Rev Lett {\bf 48} 837-8 (1982)}

{\small 7. A. Khare and T. Pradhan,(1982) Phy Rev Lett {\bf 49} 1227-8}

{\small ------(1982) Phy Rev Lett {\bf 49} 1594}

{\small ------(1983) Phy Rev Lett {\bf 51} 1108}

{\small 8. K. R. Brownstein, J. Phys A: Math Gen {\bf 19} 159-160 (1986)}

{\small 9. A.F. Ranada, J. Phys A. Math Gen.{\bf \ 25} 1621-1641 (1992)}

{\small 10. A picture of Hopf cyclide may be found at http://www.uh.edu/%
\symbol{126}rkiehn/car/carhomep.htm}

{\small 11. \ A Maple symbolic mathematics program to compute the functions
in this article may be found at }

{\small http://www.uh.edu/\symbol{126}rkiehn/maple/cyclide1.zip}

\vspace{1pt}

\vspace{1pt}

\end{document}